\documentclass[a4paper,12pt]{article}

\usepackage{enumerate}
\usepackage{graphicx}
\usepackage{amsmath}
\usepackage{amssymb}
\usepackage{hyperref}
\usepackage{tabularx}
\newcolumntype{M}{>{$}c<{$}}
\usepackage[matrix,arrow,curve]{xy}

\numberwithin{equation}{section} \numberwithin{figure}{section}
\numberwithin{table}{section}

\def\papertitlepage{\baselineskip 3.5ex\thispagestyle{empty}}
\def\Title#1{\baselineskip 1cm \vspace{1.5cm}%
  \begin{center}{\Large\bf #1}\end{center}\vspace{0.5cm}}
\def\Authors#1{\begin{center}\renewcommand{\thefootnote}{\fnsymbol{footnote}}{\it #1}\end{center}}
\def\Abstract{\vspace{1.0cm}%
  \begin{center}{\large\bf Abstract}\end{center}}

\renewenvironment{thebibliography}{\pagebreak[3]\par\vspace{0.6em}
\begin{flushleft}{\large \bf References}\end{flushleft}
\vspace{-1.0em}

\begin{enumerate}\if@twocolumn\baselineskip=0.6em\itemsep -0.2em
\else\itemsep -0.2em\fi\labelsep 0.1em}{\end{enumerate} }

\begin{document}
{\papertitlepage \vspace*{0cm} {\hfill
\begin{minipage}{4.2cm}
CCNH-UFABC 2014\par\noindent October, 2014
\end{minipage}}
\Title{Level truncation analysis of a simple tachyon vacuum
solution in cubic superstring field theory}
\Authors{{\sc E.~Aldo~Arroyo${}$\footnote{\tt
aldo.arroyo@ufabc.edu.br}}
\\
Centro de Ci\^{e}ncias Naturais e Humanas, Universidade Federal do ABC \\[-2ex]
Santo Andr\'{e}, 09210-170 S\~{a}o Paulo, SP, Brazil ${}$ }
} 

\vskip-\baselineskip
{\baselineskip .5cm \Abstract

We evaluate the vacuum energy of a simple tachyon vacuum solution
using the level truncation scheme in cubic superstring field
theory. By truncating the standard Virasoro $L_0$ level expansion
of the solution, we obtain a value of the vacuum energy in
agreement with Sen's first conjecture.
 }
\newpage
\setcounter{footnote}{0}
\tableofcontents

\section{Introduction}
The analytic solution for tachyon condensation
\cite{Schnabl:2005gv} in Witten's open bosonic string field theory
\cite{Witten:1985cc} has provided analytical and numerical tools
to analyze several classical solutions of the string field
equations of motion \cite{Okawa:2006vm}-\cite{Takahashi:2007du}.
In the framework of the modified cubic superstring field theory
\cite{Arefeva:1989cp}, the analytic construction of the tachyon
vacuum solution was analyzed by Erler \cite{Erler:2007xt}. Using
the $KBc\gamma$ subalgebra introduced in
\cite{Arroyo:2010fq,Erler:2010pr}, many gauge equivalent tachyon
vacuum solutions
\cite{Aref'eva:2008ad,Gorbachev:2010zz,Aref'eva:2009sj,Arefeva:2010yd,Kroyter:2009bg}
have been proposed and the computation of the energy associated to
these solutions was performed giving results in agreement with
Sen's conjecture. The analysis of multibrane solutions has been
given in a set of two papers
\cite{Arroyo:2013pha,AldoArroyo:2012if}. In the case of Berkovits
non-polynomial open superstring field theory
\cite{Berkovits:1995ab}, the analytic construction of the tachyon
vacuum solution, based on an extension of the $KBc\gamma$
subalgebra, has been studied by Erler \cite{Erler:2013wda}.

In the bosonic context, the well known solutions, i.e., Schnabl's
\cite{Schnabl:2005gv} and the simple Erler-Schnabl's analytic
solutions for tachyon condensation \cite{Erler:2009uj} were used
to analytically test Sen's conjecture. Apart from these analytical
computations, further numerical evidence has been provided by the
so-called level truncation analysis
\cite{Erler:2009uj,Takahashi:2007du}. In the superstring case,
only analytical calculations have been performed with the
solutions \cite{Erler:2007xt,Gorbachev:2010zz,Erler:2013wda} and
the results were in agreement with Sen's conjecture. So that the
tachyon vacuum solution from an analytic perspective appears to be
as regular as Schnabl's original solution for the bosonic string.
Nevertheless, from the perspective of the level expansion the
situation is unknown, because the analysis of the energy for the
tachyon vacuum solution in cubic superstring field theory using
the usual $L_0$ level expansion has not yet been carried out.

In this paper, we will analyze a simple analytic solution for
tachyon condensation in cubic superstring field theory. This
solution is written in terms of the elements of the $KBc\gamma$
subalgebra
\begin{eqnarray}
\label{Psisolintro} \Psi_\lambda = \frac{1}{\sqrt{1+\lambda K}}
\Big[ \frac{1}{\lambda} c +c KB c + B \gamma^2 \Big]
\frac{1}{\sqrt{1+\lambda K}} .
\end{eqnarray}
The solution (\ref{Psisolintro}) with $\lambda = 1$ has been
studied by Gorbachev \cite{Gorbachev:2010zz}, and his main results
were: (i) the analytical computation of the energy leads to a
value in agreement with Sen's first conjecture, and (ii) the proof
of the absence of physical excitations in the vicinity of the
tachyon vacuum. Let us point out that to obtain the result (i),
the validity of the equation of motion contracted with the
solution itself was assumed. However, to explicitly test the
validity of this assumption, the cubic term of the action must be
computed. By employing the solution (\ref{Psisolintro}) with a
generic $\lambda$, we will compute the kinetic $\langle
\Psi_\lambda Q \Psi_\lambda \rangle$ and the cubic terms $\langle
\Psi_\lambda \Psi_\lambda \Psi_\lambda \rangle$, and test the
validity of the equation of motion contracted with the solution
itself, namely $\langle \Psi_\lambda Q \Psi_\lambda \rangle +
\langle \Psi_\lambda \Psi_\lambda \Psi_\lambda \rangle = 0$.
Additionally, we will evaluate Ellwood's gauge invariant
observable.

Apart from the evaluation of the cubic term and Ellwood's gauge
invariant observable, we will study the solution using the
traditional Virasoro $L_0$ level truncation scheme
\cite{Kostelecky:1988ta,Kostelecky:1989nt,Moeller:2000xv,
Taylor:2002fy,Gaiotto:2002wy}. This analysis is important since we
want to know if the solution behaves as a regular element in the
state space constructed out of Fock states. Specifically the
analysis of the coefficients appearing in the $L_0$ level
expansion provides one criterion for the solution being well
defined \cite{Schnabl:2010tb,Takahashi:2007du}. Furthermore the
$L_0$ level expansion of the solution will provide us with an
additional way to numerically test Sen's first conjecture
\cite{Takahashi:2007du,Ohmori:2003vq,Raeymaekers,DeSmet:2000dp}.

The main motivation for developing a level truncation analysis of
a simple tachyon vacuum solution in cubic superstring field theory
is to prepare a numerical background in order to analyze more
cumbersome solutions, such as the multibrane solutions
\cite{Arroyo:2013pha,AldoArroyo:2012if}, and the recently proposed
Erler's analytic solution for tachyon condensation in Berkovits
non-polynomial open superstring field theory \cite{Erler:2013wda}.
Since the algebraic structure of Berkovits theory is similar to
the cubic superstring field theory, the results of our work can be
naturally extended, however the presence of a non-polynomial
action in Berkovits theory, will bring us challenges in the level
truncation analysis of Erler's solution.

This paper is organized as follows. In section 2, we review the
construction of the simple tachyon vacuum solution in cubic
superstring field theory and compute the kinetic and the cubic
terms of the action. Ellwood's gauge invariant observable is also
evaluated. In section 3, we present the level expansion analysis
of the solution. By truncating the standard Virasoro $L_0$ level
expansion of the solution, we compute the vacuum energy and obtain
a result in agreement with Sen's first conjecture. In section 4, a
summary and further directions of exploration are given. Some
details regarding the level truncation evaluation of the vacuum
energy are given in the appendix A.

\section{A simple analytic solution for tachyon condensation in cubic superstring
field theory}

In this section, we are going to review the construction of the
simple tachyon vacuum solution in cubic superstring field theory
\cite{Arroyo:2010fq}. Let us start by writing a string field
$\Psi$ as a pure gauge form
\begin{eqnarray}
\label{Solution1} \Psi = \mathcal{U} Q \mathcal{U}^{-1},
\end{eqnarray}
so that $\Psi$ formally satisfies the string field equation of
motion $Q \Psi + \Psi \Psi =0$, where $Q$ is the BRST operator of
the open Neveu-Schwarz superstring theory.

We can construct a string field $\mathcal{U}$ by employing the
elements of the $KBc\gamma$ subalgebra. The basic string fields of
this subalgebra are given by
\cite{Arroyo:2010fq,Erler:2007xt,Erler:2010pr}
\begin{eqnarray}
\label{Kdef} K &=& \frac{1}{2} \hat{\mathcal{L}} U_{1}^\dag U_{1}
|0\rangle \, ,
\\
\label{Bdef} B &=& \frac{1}{2} \hat{\mathcal{B}} U_{1}^\dag U_{1}
|0\rangle \, ,
\\
\label{cdef} c &=&   U_{1}^\dag U_{1} \tilde c (0)|0\rangle \, ,
\\
\label{gammadef} \gamma &=&   U_{1}^\dag U_{1} \tilde \gamma
(0)|0\rangle \, ,
\end{eqnarray}
where the operator $U_{1}^\dag U_{1}$ is defined in general as
$U^\dag_r U_r = e^{\frac{2-r}{2} \hat{\mathcal{L}}}$. The
operators $\hat{\mathcal{L}}$, $\hat{\mathcal{B}}$, $\tilde c(0)$
and $\tilde \gamma (0)$ are defined in the sliver frame
\footnote{Remember that a point in the upper half plane $z$ is
mapped to a point in the sliver frame $\tilde z$ via the conformal
mapping $\tilde z=\frac{2}{\pi} \arctan z $. Note that we are
using the convention of \cite{Erler:2009uj} for the conformal
mapping.}, and they are related to the worldsheet energy-momentum
tensor, the $b$, $c$ and $\gamma$ ghosts fields respectively. For
instance, the operators $\hat{\mathcal{L}}$ and
$\hat{\mathcal{B}}$ are given by
\begin{eqnarray}
\label{Lhatdef} \hat{\mathcal{L}} &\equiv& \mathcal{L}_{0} +
\mathcal{L}^{\dag}_0 = \oint \frac{d z}{2 \pi i} (1+z^{2})
(\arctan z+\text{arccot} z) \,
T(z) \, , \\
\label{Bhatdef} \hat{\mathcal{B}} &\equiv& \mathcal{B}_{0} +
\mathcal{B}^{\dag}_0 = \oint \frac{d z}{2 \pi i} (1+z^{2})
(\arctan z+\text{arccot} z) \, b(z) \, .
\end{eqnarray}

Using these definitions, we can show that the basic elements of
the $KBc\gamma$ subalgebra (\ref{Kdef})-(\ref{gammadef}) satisfy
the algebraic relations
\begin{align}
\label{relkbc1} &\{B,c\}=1\, , \;\;\;\;\;\;\; [B,K]=0 \, ,
\;\;\;\;\;\;\; B^2=c^2=0
\, ,\\
\label{relkbc2} \partial c = [K&,c] \, , \;\;\;\;\;\;\;
\partial \gamma  = [K,\gamma] \, , \;\;\;\;\;\;\; [c,\gamma]=0 \, ,
\;\;\;\;\;\;\; [B,\gamma]=0 \, ,
\end{align}
and have the following BRST variations
\begin{eqnarray}
\label{relkbc3} QK=0 \, , \;\;\;\;\;\; QB=K \, , \;\;\;\;\;\;
Qc=cKc-\gamma^2 \, , \;\;\;\;\;\; Q\gamma=c \partial \gamma
-\frac{1}{2} \gamma
\partial c \, .
\end{eqnarray}

Employing the elements of the $KBc\gamma$ subalgebra, let us write
the string field $\mathcal{U}$ as follows
\begin{eqnarray}
\label{Udef} \mathcal{U} = 1-F Bc F \;\; , \;\;\;\;
\mathcal{U}^{-1} = 1+ \frac{F}{1-F^2} Bc F,
\end{eqnarray}
where $F$ is a function of $K$ given by
\begin{eqnarray}
\label{FK} F(K) = \frac{1}{\sqrt{1+\lambda K}}.
\end{eqnarray}
Using this string field $\mathcal{U}$ and the relations
(\ref{relkbc1})-(\ref{relkbc3}), we can derive the following
solution
\begin{eqnarray}
\label{Psisol} \Psi_\lambda = \mathcal{U} Q \mathcal{U}^{-1} =
\frac{1}{\sqrt{1+\lambda K}} \Big[ \frac{1}{\lambda} c +c KB c + B
\gamma^2 \Big] \frac{1}{\sqrt{1+\lambda K}} .
\end{eqnarray}

The solution (\ref{Psisol}) with $\lambda = 1$ has been analyzed
by Gorbachev \cite{Gorbachev:2010zz}, and his main results were:
(i) the analytical computation of the energy leads to a value in
agreement with Sen's first conjecture, and (ii) the proof of the
absence of physical excitations in the vicinity of the tachyon
vacuum. Let us point out that to obtain the result (i), the
validity of the equation of motion contracted with the solution
itself was assumed. To explicitly test the validity of this
assumption, the cubic term of the action must be computed. Apart
from the evaluation of the cubic term, it remains to analyze the
solution using the traditional Virasoro $L_0$ level truncation
scheme. This analysis is important since we want to know if the
solution behaves as a regular element in the state space
constructed out of Fock states.

In the next subsections, by employing the solution (\ref{Psisol})
with a generic $\lambda$, we are going to compute the kinetic term
$\langle \Psi_\lambda Q \Psi_\lambda \rangle$ and the cubic term
$\langle \Psi_\lambda \Psi_\lambda \Psi_\lambda \rangle$, and we
will test the validity of the equation of motion contracted with
the solution itself, namely $\langle \Psi_\lambda Q \Psi_\lambda
\rangle + \langle \Psi_\lambda \Psi_\lambda \Psi_\lambda \rangle =
0$. Additionally, we will evaluate Ellwood's gauge invariant
observable. However, the main result of our paper will be
presented in section 3, i.e., the level truncation analysis of the
solution.

\subsection{Computation of the kinetic term}
In this subsection, we are going to compute the kinetic term of
the action. Since we have that $Q[Bc]= c KB c + B \gamma^2$, to
simplify the computation it will be useful to write the solution
(\ref{Psisol}) in the following way
\begin{eqnarray}
\label{Psisolbrst} \Psi_\lambda = \frac{1}{\sqrt{1+\lambda K}}
\frac{c}{\lambda} \frac{1}{\sqrt{1+\lambda K}} + Q\Big\{
\frac{1}{\sqrt{1+\lambda K}} Bc \frac{1}{\sqrt{1+\lambda K}}
\Big\}.
\end{eqnarray}
Plugging this solution into the definition of the kinetic term
$\langle \Psi_\lambda Q \Psi_\lambda \rangle$, we get
\begin{eqnarray}
\label{kineticcom1} \langle \Psi_\lambda Q \Psi_\lambda \rangle =
\frac{1}{\lambda^2} \big\langle c \frac{1}{1+\lambda K} Qc
\frac{1}{1+\lambda K} \big\rangle.
\end{eqnarray}

The action of the BRST operator on the $c$ field is $Qc =
cKc-\gamma^2$. Since the non-vanishing correlators for elements in
the $KBc\gamma$ subalgebra, in cubic superstring field theory, are
proportional to $\langle c F_1(K)\gamma^2 F_2(K) \rangle$ or
$\langle B \gamma^2 F_1(K) c F_2(K) c F_3(K) \rangle$, the
non-vanishing contribution to the kinetic term (\ref{kineticcom1})
is given by
\begin{eqnarray}
\label{kineticcom2} \langle \Psi_\lambda Q \Psi_\lambda \rangle =
- \frac{1}{\lambda^2} \big\langle c \frac{1}{1+\lambda K} \gamma^2
\frac{1}{1+\lambda K} \big\rangle.
\end{eqnarray}
Let us write the integral representation of the function
$1/(1+\lambda K)$,
\begin{eqnarray}
\label{integra1} \frac{1}{1+\lambda K} = \int_0^{\infty} dt \,
e^{-t (1+\lambda K)} = \int_0^{\infty} dt \, e^{-t}
\Omega^{\lambda t},
\end{eqnarray}
where $\Omega= e^{-K}$. Plugging this integral representation into
the equation (\ref{kineticcom2}), we obtain
\begin{eqnarray}
\label{kineticcom3} \langle \Psi_\lambda Q \Psi_\lambda \rangle =
- \frac{1}{\lambda^2} \int_0^{\infty} dt_1 dt_2 \,
e^{-t_1-t_2}\big\langle c \Omega^{\lambda t_1} \gamma^2
\Omega^{\lambda t_2} \big\rangle,
\end{eqnarray}
where the correlator $\langle c \Omega^{\lambda t_1} \gamma^2
\Omega^{\lambda t_2} \rangle$ is given by \cite{Erler:2007xt,
Arroyo:2010fq}
\begin{eqnarray}
\label{corre1x1} \langle c \Omega^{\lambda t_1} \gamma^2
\Omega^{\lambda t_2} \rangle = \frac{ \lambda^2(t_1+t_2)^2}{2
\pi^2},
\end{eqnarray}
so that plugging equation (\ref{corre1x1}) into equation
(\ref{kineticcom3}), we get
\begin{eqnarray}
\label{kinec3} \langle \Psi_\lambda Q \Psi_\lambda \rangle
=-\frac{1}{2 \pi^2 } \int_{0}^{\infty} dt_1  dt_2 \, e^{-t_1-t_2}
(t_1+t_2)^2 ,
\end{eqnarray}
where the integral has the value
\begin{eqnarray}
\label{integra1} \int_{0}^{\infty} dt_1  dt_2 \, e^{-t_1-t_2}
(t_1+t_2)^2 = 6 ,
\end{eqnarray}
therefore the value of the kinetic term is
\begin{eqnarray}
\label{kinec4} \langle \Psi_\lambda Q \Psi_\lambda \rangle
=-\frac{3}{ \pi^2 },
\end{eqnarray}
as expected.

\subsection{Computation of the cubic term}
In this subsection, we are going to compute the cubic term of the
action. Plugging the solution (\ref{Psisolbrst}) into the
definition of the cubic term $ \langle \Psi_\lambda \Psi_\lambda
\Psi_\lambda \rangle $, and using the fact that the only
non-vanishing correlators are proportional to $\langle c
F_1(K)\gamma^2 F_2(K) \rangle$ or $\langle B \gamma^2 F_1(K) c
F_2(K) c F_3(K) \rangle$, after performing some algebraic
manipulations, we obtain
\begin{align}
\label{cubic1} \langle \Psi_\lambda \Psi_\lambda \Psi_\lambda
\rangle &= \frac{3}{\lambda^2} \big\langle B \gamma^2
\frac{1}{1+\lambda K} c \frac{1}{1+\lambda K} c \frac{1}{1+\lambda
K} \big\rangle + \frac{3}{\lambda} \big\langle B \gamma^2
\frac{1}{1+\lambda K} c \frac{1}{1+\lambda K} c \frac{K}{1+\lambda
K} \big\rangle \nonumber \\
&+ \frac{3}{\lambda} \big\langle B \gamma^2 \frac{K}{1+\lambda K}
c \frac{1}{1+\lambda K} c \frac{1}{1+\lambda K} \big\rangle.
\end{align}
Let us write the integral representation for the corresponding
functions,
\begin{eqnarray}
\label{integray1} \frac{1}{1+\lambda K} &=& \int_0^{\infty} dt \,
e^{-t (1+\lambda K)} = \int_0^{\infty} dt \, e^{-t}
\Omega^{\lambda t}, \\
\label{integray2}\frac{K}{1+\lambda K} &=&
-\frac{1}{\lambda}\int_0^{\infty} dt \, e^{-t}
\frac{\partial}{\partial t}(e^{-t \lambda K}) = -\frac{1}{\lambda}
\int_0^{\infty} dt \, e^{-t} \frac{\partial}{\partial
t}\Omega^{\lambda t}.
\end{eqnarray}
Plugging these integral representations into the equation
(\ref{cubic1}), we get
\begin{align}
\label{cubic2} \langle \Psi_\lambda \Psi_\lambda \Psi_\lambda
\rangle &= \frac{3}{\lambda^2} \int_{0}^{\infty} dt_1  dt_2 dt_3\,
e^{-t_1-t_2-t_3} \Big( 1-\frac{\partial}{\partial t_1} -
\frac{\partial}{\partial t_3}\Big) \langle B \gamma^2
\Omega^{\lambda t_1} c \Omega^{\lambda t_2}  c \Omega^{\lambda
t_3} \rangle ,
\end{align}
where the correlator $\langle B \gamma^2 \Omega^{\lambda t_1} c
\Omega^{\lambda t_2}  c \Omega^{\lambda t_3} \rangle$ is given by
\cite{Erler:2007xt, Arroyo:2010fq}
\begin{eqnarray}
\label{corre1y2} \langle B \gamma^2 \Omega^{\lambda t_1} c
\Omega^{\lambda t_2}  c \Omega^{\lambda t_3} \rangle = \frac{
\lambda^2(t_1+t_2+t_3)t_2}{2 \pi^2},
\end{eqnarray}
so that plugging equation (\ref{corre1y2}) into equation
(\ref{cubic2}), we get
\begin{eqnarray}
\label{cubic3} \langle \Psi_\lambda \Psi_\lambda \Psi_\lambda
\rangle =\frac{3}{2 \pi^2 } \int_{0}^{\infty} dt_1  dt_2 dt_3\,
e^{-t_1-t_2-t_3} (t_1+t_2+t_3-2)t_2 ,
\end{eqnarray}
where the integral has the value
\begin{eqnarray}
\label{integracub1} \int_{0}^{\infty} dt_1  dt_2  dt_3 \,
e^{-t_1-t_2-t_3} (t_1+t_2+t_3-2)t_2 = 2,
\end{eqnarray}
therefore the value of the cubic term is
\begin{eqnarray}
\label{cubic4} \langle \Psi_\lambda \Psi_\lambda \Psi_\lambda
\rangle= +\frac{3}{ \pi^2 },
\end{eqnarray}
as expected.

Employing the result of the kinetic term (\ref{kinec4}) and the
cubic term (\ref{cubic4}), we just have proven the validity of the
equation of motion contracted with the solutions itself, namely:
$\langle \Psi_\lambda Q \Psi_\lambda \rangle + \langle
\Psi_\lambda \Psi_\lambda \Psi_\lambda \rangle = 0$. In the next
subsection, we are going to evaluate another important gauge
invariant quantity.

\subsection{Computation of Ellwood's gauge invariant}
Let us evaluate Ellwood's gauge invariant overlap for the tachyon
vacuum solution. For a generic solution $\Psi$, Ellwood's gauge
invariant overlap is defined as
\begin{eqnarray}
\label{wood1} W(\Psi,\mathcal{V}) = \text{Tr}(\Psi),
\end{eqnarray}
where the notation Tr$(\cdots)$ refers to a correlator with an on
shell closed string vertex operator $\mathcal{V}( i )$ inserted at
the midpoint, $\text{Tr}(\Psi)=\langle \mathcal{V}( i ) \Psi
\rangle$. We assume the same $\mathcal{V}$ used in reference
\cite{Erler:2010pr}, this field is an NS-NS closed string vertex
operator of the form
\begin{eqnarray}
\label{wood2} \mathcal{V}(z)= c \tilde{c} e^{-\phi}
e^{-\tilde{\phi}} \mathcal{O}^{\text{m}},
\end{eqnarray}
where $\mathcal{O}^{\text{m}}$ is a weight $(\frac{1}{2},
\frac{1}{2})$ superconformal matter primary field. As argued by
Ellwood \cite{Ellwood:2008jh}, the gauge invariant overlap
represents the shift in the closed string tadpole of the solution
relative to the perturbative vacuum.

Inserting the solution (\ref{Psisolbrst}) into the definition of
the gauge invariant overlap (\ref{wood1}), the BRST exact term
does not contribute, therefore we get
\begin{eqnarray}
\label{wood3} W(\Psi_\lambda,\mathcal{V}) = \frac{1}{\lambda}
\text{Tr}(\frac{1}{\sqrt{1+\lambda K}} c \frac{1}{\sqrt{1+\lambda
K}}).
\end{eqnarray}
Let us write the integral representation of the function
$1/\sqrt{1+\lambda K}$,
\begin{eqnarray}
\label{integrawood1} \frac{1}{\sqrt{1+\lambda K}} =
\frac{1}{\sqrt{\pi}} \int_0^{\infty} dt \, \frac{1}{\sqrt{t}}
e^{-t (1+\lambda K)} = \frac{1}{\sqrt{\pi}} \int_0^{\infty} dt \,
\frac{1}{\sqrt{t}} e^{-t} \Omega^{\lambda t}.
\end{eqnarray}
Plugging this integral representation into the equation
(\ref{wood3}), we obtain
\begin{eqnarray}
\label{woodcom1} W(\Psi_\lambda,\mathcal{V}) = \frac{1}{\lambda
\pi} \int_0^{\infty} dt_1 dt_2 \, \frac{1}{\sqrt{t_1 t_2}}
e^{-t_1-t_2} \text{Tr}(\Omega^{\lambda t_1} c \Omega^{\lambda t_2}
).
\end{eqnarray}
The correlator $\text{Tr}(\Omega^{\lambda t_1} c \Omega^{\lambda
t_2} )$ is given by \cite{Erler:2010pr,Arroyo:2013pha}
\begin{eqnarray}
\label{correwoodx1} \text{Tr}(\Omega^{\lambda t_1} c
\Omega^{\lambda t_2} ) = \lambda (t_1+t_2) \text{Tr}( c \Omega) ,
\end{eqnarray}
where $\Omega = e^{-K}$ and $\text{Tr}( c \Omega) = \langle
\mathcal{V}(i \infty) c(0)\rangle_{C_{1}}$ is the expected result
of the closed string tadpole on the disk. Plugging the correlator
(\ref{correwoodx1}) into equation (\ref{woodcom1}), we get
\begin{eqnarray}
\label{woodcom2} W(\Psi_\lambda,\mathcal{V}) = \frac{1}{\pi}
\int_0^{\infty} dt_1 dt_2 \, \frac{1}{\sqrt{t_1 t_2}} e^{-t_1-t_2}
(t_1+t_2) \text{Tr}( c \Omega) ,
\end{eqnarray}
where the integral has the value
\begin{eqnarray}
\label{integrawoody1} \int_0^{\infty} dt_1 dt_2 \,
\frac{1}{\sqrt{t_1 t_2}} e^{-t_1-t_2} (t_1+t_2) = \pi ,
\end{eqnarray}
therefore the value of Ellwood's gauge invariant is
\begin{eqnarray}
\label{woodcom3} W(\Psi_\lambda,\mathcal{V}) =(+1) \text{Tr}( c
\Omega) ,
\end{eqnarray}
as expected.

Since the analytic computation of the gauge invariant quantities
performed in the previous subsections leads to the desired
results, the tachyon vacuum solution (\ref{Psisol}) from an
analytic perspective appears to be as regular as Schnabl's
original solution for the bosonic string. Let us see what happens
from a numerical perspective. In what follows, we are going to
analyze the solution using the usual $L_0$ level expansion scheme.

\section{Level expansion analysis}
In this section, we are going to analyze the level expansion of
the simple tachyon vacuum solution (\ref{Psisolbrst}). The
analysis of a string field using the traditional $L_0$ level
expansion scheme is important since this information tells us if
the solution behaves as a regular element in the state space
constructed out of Fock states. Specifically the analysis of the
coefficients appearing in the $L_0$ level expansion provides one
criterion for the solution being well defined
\cite{Schnabl:2010tb,Takahashi:2007du}. Moreover the $L_0$ level
expansion of the solution brings an additional way to numerically
test Sen's first conjecture.

\subsection{$L_0$ level expansion of the simple tachyon vacuum
solution}

To expand the simple tachyon vacuum solution (\ref{Psisolbrst}) in
the Virasoro basis of $L_0$ eigenstates, we start by writing the
function $1/\sqrt{1+\lambda K}$ as its integral representation
(\ref{integrawood1}). Plugging this integral representation
(\ref{integrawood1}) into the expression for the simple tachyon
vacuum solution (\ref{Psisolbrst}), we obtain
\begin{eqnarray}
\label{LevelPx1} \Psi_{\lambda} = \frac{1}{\pi \lambda }
\int_{0}^{\infty}ds dt \, \frac{1}{\sqrt{st}} e^{-s-t}
\Omega^{\lambda t} c \Omega^{\lambda s}   + Q \Big\{ \frac{1}{\pi
} \int_{0}^{\infty}ds dt \, \frac{1}{\sqrt{st}} e^{-s-t}
\Omega^{\lambda t} Bc \Omega^{\lambda s}  \Big\} ,
\end{eqnarray}
where the wedge state $\Omega^{t}$ can be expressed in terms of
the scaling operator $U_r$ \cite{Erler:2009uj,Schnabl:2002gg}
\begin{eqnarray}
\label{omePx1} \Omega^t = e^{-t K} = U^{\dagger}_{t+1} U_{t+1}| 0
\rangle, \;\;\;\;\;\text{where}\;\;\;\;\; U_{r} \equiv
\Big(\frac{2}{r} \Big)^{\mathcal{L}_0}.
\end{eqnarray}

Let us write rather general formulas which will be very useful for
the $L_0$ level expansion analysis of a string field constructed
out of elements in the $KBc\gamma$ subalgebra,
\begin{eqnarray}
\label{ideng35yut} e^{-t_1 K} c e^{-t_2K} B c e^{-t_3K} = \frac{r
\cos ^2\left(\frac{\pi x}{r}\right) \left(\pi  (r-2 y)-r \sin
\left(\frac{2 \pi y}{r}\right)\right)}{4 \pi^2 }\widetilde{U}_{r}
c\left(\frac{2 \tan \left(\frac{\pi
   x}{r}\right)}{r}\right)|0\rangle \;\;\; \nonumber \\
   +  \frac{r \cos ^2\left(\frac{\pi  y}{r}\right) \left(\pi  (r+2 x)+r \sin \left(\frac{2 \pi  x}{r}\right)\right)}{4 \pi^2 }
   \widetilde{U}_{r} c\left(\frac{2 \tan
   \left(\frac{\pi
   y}{r}\right)}{r}\right)|0\rangle \;\;\; \nonumber \\
+ \sum_{k=1}^{\infty}\frac{(-1)^{k+1} 2^{2 k-1}
\left(\frac{1}{r}\right)^{2 k-3} \cos ^2\left(\frac{\pi
x}{r}\right) \cos ^2\left(\frac{\pi  y}{r}\right)}{\left(4
k^2-1\right)
   \pi^2 } \widetilde{U}_{r}
   b_{-2k} c\left(\frac{2 \tan \left(\frac{\pi
   x}{r}\right)}{r}\right) c\left(\frac{2 \tan \left(\frac{\pi
   y}{r}\right)}{r}\right)|0\rangle , \; \\ r=t_1+t_2+t_3+1 , \;\;\;\;\;
x=\frac{1}{2}(t_3-t_1-t_2), \;\;\;\;\; y=\frac{1}{2}(t_2+t_3-t_1), \;\;\;\;\;\;\;\; \\
e^{-t K} B \gamma^2 e^{-t K} =  \frac{2}{\pi} \sum_{k=1}^{\infty}
\widetilde{U}_{2t+1} \frac{
(-1)^{k+1}}{4k^2-1}\big(\frac{2}{2t+1}\big)^{2k-1} b_{-2k}\gamma^2
(0)|0\rangle, \;\;\;\;\;\;\;\;\;
\end{eqnarray}
where the operator $\widetilde{U}_{r}$ is defined as
\begin{eqnarray}
\label{uux2} \widetilde{U}_{r} \equiv  \cdots e^{u_{10,r} L_{-10}}
e^{u_{8,r} L_{-8}} e^{u_{6,r} L_{-6}} e^{u_{4,r} L_{-4}}e^{u_{2,r}
L_{-2}}.
\end{eqnarray}
To find the coefficients $u_{n,r}$ appearing in the exponentials,
we use
\begin{align}
\frac{r}{2} \tan (\frac{2}{r} \arctan z) &= \lim_{N \rightarrow
\infty} \big[f_{2,u_{2,r}} \circ  f_{4,u_{4,r}} \circ
f_{6,u_{6,r}} \circ f_{8,u_{8,r}} \circ
f_{10,u_{10,r}} \circ \cdots \circ f_{N,u_{N,r}}(z)\big] \nonumber \\
&= \lim_{N \rightarrow \infty}\big[ f_{2,u_{2,r}} ( f_{4,u_{4,r}}
( f_{6,u_{6,r}} ( f_{8,u_{8,r}} ( f_{10,u_{10,r}}(\cdots
(f_{N,u_{N,r}}(z)) \dots )))))  \big],
\end{align}
where the function $f_{n,u_{n,r}}(z)$ is given by
\begin{eqnarray}
f_{n,u_{n,r}}(z) = \frac{z}{(1-u_{n,r} n z^n)^{1/n}}.
\end{eqnarray}

Employing the set of equations (\ref{omePx1})$-$(\ref{uux2}) for
the solution (\ref{LevelPx1}), we obtain
\begin{eqnarray}
\label{GSollevel} \Psi_{\lambda} = \frac{1}{2 \pi^2 \lambda }
\int_{0}^{\infty}ds dt \, \frac{1}{\sqrt{st}} e^{-s-t} r^2 \cos
^2\left(\frac{\pi x}{r}\right) \widetilde{U}_{r} c\left(\frac{2
\tan \left(\frac{\pi
   x}{r}\right)}{r}\right)|0\rangle   + Q_{\text{exact term}} ,
\end{eqnarray}
where
\begin{eqnarray}
\label{rx} r=\lambda (s+t)+1 , \;\;\;\;\;
x=\frac{\lambda}{2}(s-t).
\end{eqnarray}
Since in the evaluation of the vacuum energy, the $Q_{\text{exact
term}}$ will not contribute, we only need to consider the first
term on the right hand side of equation (\ref{GSollevel}). Let us
study in some detail this first term
\begin{eqnarray}
\label{GSollevelFirst1} \Psi^{(1)}_{\lambda} \equiv \frac{1}{2
\pi^2 \lambda } \int_{0}^{\infty}ds dt \, \frac{1}{\sqrt{st}}
e^{-s-t} r^2 \cos ^2\left(\frac{\pi x}{r}\right) \widetilde{U}_{r}
c\left(\frac{2 \tan \left(\frac{\pi
   x}{r}\right)}{r}\right)|0\rangle ,
\end{eqnarray}
with $r$ and $x$ given by equation (\ref{rx}).

By writing the $c$ ghost in terms of its modes $c(z)=\sum_{m}
c_m/z^{m-1}$ and employing equations (\ref{uux2}) and
(\ref{GSollevelFirst1}), we can expand $\Psi^{(1)}_{\lambda}$ in
terms of the elements contained in the Virasoro basis of $L_0$
eigenstates. For instance, let us expand $\Psi^{(1)}_\lambda$ up
to level two states
\begin{eqnarray}
\label{PsiY1} \Psi^{(1)}_\lambda= t(\lambda) c_1 | 0 \rangle +
v(\lambda) c_{-1} | 0 \rangle + w(\lambda) L_{-2}c_1 | 0 \rangle +
\cdots ,
\end{eqnarray}
where the coefficients of the expansion $t(\lambda)$, $v(\lambda)$
and $w(\lambda)$  are given by the following integrals
\begin{align}
t(\lambda) &= \int_{0}^{\infty}ds dt \, \frac{e^{-s-t} (\lambda  (s+t)+1)^2
\cos ^2\left(\frac{\pi  \lambda  (s-t)}{2 (\lambda  (s+t)+1)}\right)}{2 \pi ^2 \lambda  \sqrt{s t}}, \\
v(\lambda) &= \int_{0}^{\infty}ds dt \, \frac{2 e^{-s-t} \sin ^2\left(\frac{\pi  \lambda  (s-t)}{2 (\lambda  (s+t)+1)}\right)}{\pi ^2 \lambda  \sqrt{s t}}, \\
w(\lambda) &= \int_{0}^{\infty}ds dt \, \frac{e^{-s-t}
\left(4-(\lambda  (s+t)+1)^2\right) \cos ^2\left(\frac{\pi
\lambda  (s-t)}{2 (\lambda  (s+t)+1)}\right)}{6 \pi ^2 \lambda
\sqrt{s
   t}} .
\end{align}
These integrals are convergent provided that the parameter
$\lambda$ belongs to the interval $(0,+\infty)$. By performing the
change of variables
\begin{eqnarray}
s\to \frac{1}{2} (u-u \eta), \;\;\;\; t\to \frac{1}{2} (u+u \eta)
, \;\;\;\; ds dt \to \frac{u}{2} dud\eta,
\end{eqnarray}
where $u \in[0,\infty)$ and $\eta \in (-1,1)$, and employing
numerical values for the parameter $\lambda$, we are going to
numerically evaluate the integrals.

\subsection{Level truncation evaluation of the vacuum energy}
Since in subsection 2.2 we have shown the validity of the equation
of motion contracted with the solution itself, we can write the
following expression for the normalized value of the vacuum energy
\begin{eqnarray}
\label{enernormalized1} E_\lambda = \frac{\pi^2}{3} \langle
\Psi_\lambda, Q \Psi_\lambda\rangle.
\end{eqnarray}

Plugging the simple tachyon vacuum solution (\ref{GSollevel}) into
equation (\ref{enernormalized1}), and using the fact that the
second term on the right hand side of equation (\ref{GSollevel})
does not contribute since it is a BRST exact term, we obtain
\begin{eqnarray}
\label{enernormalized2} E_\lambda = \frac{\pi^2}{3} \langle
\Psi^{(1)}_\lambda, Q \Psi^{(1)}_\lambda\rangle,
\end{eqnarray}
where the string field $\Psi^{(1)}_\lambda$ is defined in equation
(\ref{GSollevelFirst1}). As described in the bosonic case
\cite{Schnabl:2005gv,Erler:2009uj,Arroyo:2009ec,Takahashi:2007du},
it is convenient to replace the string field $\Psi^{(1)}_\lambda$
with $z^{L_0} \Psi^{(1)}_\lambda$ in the $L_0$ level truncation
scheme, so that states in the $L_0$ level expansion of the
solution acquire different integer powers of $z$ at different
levels. This parameter $z$ is needed because we will use Pad\'e
approximants to evaluate the normalized value of the vacuum energy
(\ref{enernormalized2}). After doing the calculations, we will
simply set $z = 1$.

By writing the $c$ ghost in terms of its modes and employing
equations (\ref{uux2}) and (\ref{GSollevelFirst1}), the string
field $\Psi^{(1)}_\lambda$ can be readily expanded and the
individual coefficients can be numerically integrated. As an
example, employing some specific values for the parameter
$\lambda$, let us write the string field expanded up to the level
fourth states
\begin{align}
\label{psi1over4} \Psi^{(1)}_{\lambda=1/4 } &= 0.94629\, c_1| 0
\rangle+0.138297\, c_{-1}| 0
\rangle+0.487297 \, L_{-2}c_1| 0 \rangle + 0.0574864\, c_{-3}| 0 \rangle \nonumber \\
& -0.19709 \, L_{-4}c_1| 0 \rangle+0.0390564 \,L_{-2}c_{-1}| 0
\rangle+0.166056\,L_{-2}L_{-2}c_1| 0 \rangle+\cdots,\\
\label{psi1over2} \Psi^{(1)}_{\lambda=1/2 } &= 0.635128\, c_1| 0
\rangle+0.148658 \, c_{-1}| 0
\rangle+0.163151\, L_{-2}c_1| 0 \rangle + 0.085191 \, c_{-3}| 0 \rangle \nonumber \\
& -0.06659\, L_{-4}c_1| 0 \rangle+0.0144386 \,L_{-2}c_{-1}| 0
\rangle+0.056610\,L_{-2}L_{-2}c_1| 0 \rangle+\cdots,\\
\label{psi3over4} \Psi^{(1)}_{\lambda=3/4 } &= 0.543591\, c_1| 0
\rangle+0.141941 \, c_{-1}| 0
\rangle+0.0544315\, L_{-2}c_1| 0 \rangle + 0.09104 \, c_{-3}| 0 \rangle \nonumber \\
& -0.02977\, L_{-4}c_1| 0 \rangle+0.0006467\,L_{-2}c_{-1}| 0
\rangle+0.031485\,L_{-2}L_{-2}c_1| 0 \rangle+\cdots,\\
\label{psi2over2} \Psi^{(1)}_{\lambda=1 } &= 0.509038\, c_1| 0
\rangle+0.13231\, c_{-1}| 0
\rangle-0.00157617\, L_{-2}c_1| 0 \rangle +0.08933\, c_{-3}| 0 \rangle \nonumber \\
& -0.01357\, L_{-4}c_1| 0 \rangle-0.0069469\,L_{-2}c_{-1}| 0
\rangle+0.023157\,L_{-2}L_{-2}c_1| 0 \rangle+\cdots,\\
\label{psi5over4} \Psi^{(1)}_{\lambda=5/4} &= 0.498059\, c_1| 0
\rangle+0.12281\, c_{-1}| 0
\rangle-0.0371919\, L_{-2}c_1| 0 \rangle + 0.085046\, c_{-3}| 0 \rangle \nonumber \\
& -0.00471 \, L_{-4}c_1| 0 \rangle-0.0112661\,L_{-2}c_{-1}| 0
\rangle+0.020254\,L_{-2}L_{-2}c_1| 0 \rangle+\cdots.
\end{align}

As mentioned previously, to evaluate the normalized value of the
vacuum energy (\ref{enernormalized2}), first we perform the
replacement $\Psi^{(1)}_\lambda \rightarrow z^{L_0}
\Psi^{(1)}_\lambda$ and then using the resulting string field
$z^{L_0} \Psi^{(1)}_\lambda$, we define
\begin{eqnarray}
\label{energydependz} E_\lambda(z) \equiv \frac{\pi^2}{3} \langle
z^{L_0} \Psi^{(1)}_\lambda, Q z^{L_0} \Psi^{(1)}_\lambda\rangle.
\end{eqnarray}
The normalized value of the vacuum energy (\ref{enernormalized2})
is obtained just by setting $z = 1$ in equation
(\ref{energydependz}). As we can see, our problem has been reduced
to computation of correlation functions of elements contained in
the Virasoro basis of $L_0$ eigenstates. Some details regarding
this computation are presented in appendix A. Plugging the level
expansions (\ref{psi1over4})-(\ref{psi5over4}) into the definition
(\ref{energydependz}), and after evaluating the appropriate
correlators, we obtain
\begin{eqnarray}
\label{enerz3over8} E_{\lambda=1/4}(z)
&=&-\frac{1.47298}{z^2}-0.43054+1.99584 z^2+0.21657 z^4-2.47629
z^6+\cdots, \;\;\;\;\;
\\
\label{enerz1over2} E_{\lambda=1/2}(z)
&=&-\frac{0.66354}{z^2}-0.31061+0.27892 z^2+0.03277 z^4-0.34469
z^6+\cdots, \;\;\;\;\;
\\
\label{enerz3over4} E_{\lambda=3/4}(z)
&=&-\frac{0.48606}{z^2}-0.25383+0.01408 z^2-0.01013 z^4-0.09796
z^6+\cdots, \;\;\;\;\;
\\
\label{enerz2over2} E_{\lambda=1}(z)
&=&-\frac{0.42623}{z^2}-0.22157-0.05273 z^2-0.02121 z^4-0.03456
z^6+\cdots, \;\;\;\;\;
\\
\label{enerz5over4} E_{\lambda=5/4}(z)
&=&-\frac{0.40804}{z^2}-0.20123-0.06765 z^2-0.02235 z^4-0.00978
z^6+\cdots. \;\;\;\;\;
\end{eqnarray}
As in the bosonic case \cite{Erler:2009uj,AldoArroyo:2011gx},
using these kind of series in $z$ for $E_{\lambda}(z)$, we will
compute the normalized value of the vacuum energy by the standard
procedure based on Pad\'{e} approximants. To obtain a Pad\'{e}
approximant of order $P^n_{2+n}(\lambda,z)$ for the energy, we
will need to know the series expansion of $E_{\lambda}(z)$ up to
the order $z^{2n-2}$.

For the numerical evaluation of the vacuum energy, we have
considered the string field $\Psi^{(1)}_\lambda$ expanded up to
the level twelve states, so that we obtain a series expansion for
$E_{\lambda}(z)$ truncated up to the order $z^{22}$. For instance,
the explicit expression for the vacuum energy with $\lambda=1/4$,
which includes contributions from the string field
$\Psi^{(1)}_{\lambda=1/4}$ truncated up to the level twelve
states, is given by
\begin{align}
E_{\lambda=1/4}(z) =& - \frac{1.472981}{z^2} -0.430541 +1.995842
z^2 +0.073704 z^4 -2.717863 z^6  \nonumber
\\ & -0.0485597 z^8  +4.020699 z^{10} +0.098363 z^{12} -6.0537501 z^{14} \nonumber
\\ \label{enerlevel22} & +0.033435 z^{16} +9.266269 z^{18}  +0.284437 z^{20} -12.515001 z^{22}.
\end{align}

As an illustration of the numerical method based on Pad\'{e}
approximants, let us compute the normalized value of the vacuum
energy using a Pad\'{e} approximant of order
$P^4_{2+4}(\lambda,z)$. First, we express $E_{\lambda}(z)$ as the
rational function $P^4_{2+4}(\lambda,z)$, in this example we
consider $\lambda=1/4$
\begin{eqnarray}
\label{ss2} E_{\lambda=1/4}(z)=P^4_{2+4}(1/4,z)=\frac{1}{z^2}
\Big[\frac{a_0+a_1z+a_2z^2+a_3z^3+a_4z^4
}{1+b_1z+b_2z^2+b_3z^3+b_4z^4} \Big]\, .
\end{eqnarray}
Expanding the right hand side of (\ref{ss2}) around $z=0$ up to
sixth order in $z$ and equating the coefficients of $z^{-2}$,
$z^{-1}$, $z^{0}$, $z^{1}$, $z^{2}$, $z^{3}$, $z^{4}$, $z^{5}$,
$z^{6}$ with the expansion (\ref{enerlevel22}), we get a system of
algebraic equations for the unknown coefficients $a_0$, $a_1$,
$a_2$, $a_3$, $a_4$, $b_1$, $b_2$, $b_3$ and $b_4$. Solving those
equations we get
\begin{eqnarray}
a_0 = -1.47298, \;\;\; a_1=0, \;\;\; a_2=-0.805856 ,
\;\;\; a_3=0, \;\;\; a_4=-0.10585, \;\;\;\; \\
b_1 = 0, \;\;\; b_2=0.254799, \;\;\; b_3=0, \;\;\; b_4=1.35235.
\;\;\;\;\;\;\;\;\;\;\;\;\;\;\;\;\;\;\;\;\;
\end{eqnarray}
Replacing the value of these coefficients inside the definition of
$P^4_{2+4}(1/4,z)$ (\ref{ss2}), and evaluating this at $z=1$, we
get the following normalized value of the vacuum energy
\begin{eqnarray}
\label{vacum01x} P^4_{2+4}(1/4,z=1) = -0.914671336.
\end{eqnarray}
If we naively evaluate the truncated vacuum energy
(\ref{enerlevel22}), i.e., setting $z=1$ in the series before
using Pad\'{e} approximants, we obtain a non-convergent sum. This
kind of divergence is also present in the bosonic case
\cite{Erler:2009uj}, where such behavior has appeared in the
canonical $L_0$ level truncation scheme. And therefore to
numerically evaluate the vacuum energy, the use of Pad\'{e}
approximants has been necessary.

The results of our calculations are summarized in table
\ref{realresultsx1}. As we can see, the normalized value of the
vacuum energy evaluated using Pad\'{e} approximants confirms the
expected analytic result (\ref{kinec4}). Although the convergence
to the expected answer gets quite slow, by considering higher
level contributions, we will eventually reach to the right value
of the vacuum energy $E \rightarrow -1$.

\begin{table}[ht]
\caption{The Pad\'{e} approximation for the normalized value of
the vacuum energy $\frac{\pi^2}{3} \langle z^{L_0}
\Psi^{(1)}_\lambda, Q z^{L_0} \Psi^{(1)}_\lambda\rangle$ evaluated
at $z=1$. The results show the $P_{2+n}^{n}$ Pad\'{e}
approximation for various values of the parameter $\lambda$.}
\centering
\begin{tabular}{|c|c|c|c|c|c|}
\hline
  &  $P^{n}_{2+n}[\lambda = 1/4]$   &  $P^{n}_{2+n}[\lambda = 1/2 ]$ &
  $P^{n}_{2+n}[\lambda = 3/4]$ & $P^{n}_{2+n}[\lambda = 1 ]$ & $P^{n}_{2+n}[\lambda = 5/4]$ \\
    \hline $n=0$  & $-1.47298157$ &  $-0.66354579$ &  $-0.48606372$ & $-0.42623492$ & $-0.40804682$ \\
\hline  $n=2$ & $-1.54937747$ & $-0.82720585$  & $-0.72655866$ & $-0.71701445$ & $-0.71120414$   \\
\hline $n=4$ & $-0.91467133$  & $-0.79944201$ & $-0.70410295$ & $-0.74899064$ & $-0.84724417$  \\
\hline $n=6$ & $-0.98146315$ & $-0.90224564$ & $-0.84739251$ & $-0.78062826$ & $-0.87849745$  \\
\hline $n=8$  & $-0.96055717$ &  $-1.02422931$ &  $-0.89416027$ & $-0.84178146$ & $-0.80235684$ \\
\hline  $n=10$ & $-0.98359644$ & $-0.92679142$  & $-0.89860136$ & $-0.84459366$ & $-0.78508771$   \\
\hline $n=12$  & $-0.94981557$ & $-0.93042155$ & $-1.01894842$ & $-0.85194204$ & $-0.78722918$  \\
\hline
\end{tabular}
\label{realresultsx1}
\end{table}

\section{Summary and discussion}
We have analyzed a simple tachyon vacuum solution in cubic
superstring field theory. Using this solution, we have tested the
validity of the equation of motion contracted with the solution
itself, and have evaluated Ellwood's gauge invariant observable.
However, the main result of our paper has been the level
truncation analysis of the solution. We have seen that the
solution behaves as a regular element in the state space generated
by the Virasoro basis of $L_0$ eigenstates. We have shown that the
computation of the vacuum energy using the level truncated
solution brings a value in agreement with Sen's conjecture.

Using the level truncation scheme, it would also be interesting to
analyze Schnabl type tachyon vacuum solution which was proposed by
Erler \cite{Erler:2007xt}. This solution, like the original
Schnabl's bosonic solution \cite{Schnabl:2005gv}, has an extra
term known as the phantom term, this term is necessary for the
equation of motion contracted with the solution itself to be
satisfied \cite{Okawa:2006vm}. The level truncation analysis of
Schnabl's bosonic solution has been carried out in reference
\cite{Takahashi:2007du}, and it has been shown that, to
numerically obtain the value of the vacuum energy, the use of
Pad\'{e} approximants was not necessary. It should be nice to see
if the same phenomenon can happen in the superstring context,
namely if the evaluation of the vacuum energy using the truncated
solution leads to a convergent series for the energy.

The main motivation for studying the level truncation analysis of
tachyon vacuum solutions in cubic superstring field theory is to
prepare a numerical background in order to analyze more cumbersome
solutions, such as the multibrane solutions
\cite{Arroyo:2013pha,AldoArroyo:2012if}, and the recently proposed
Erler's analytic solution for tachyon condensation in Berkovits
non-polynomial open superstring field theory \cite{Erler:2013wda}.
The presence of higher interaction terms in Berkovits string field
theory action, will bring us challenges for the level truncation
analysis of Erler's solution. In a future work, we would like to
extend the results of our paper, for instance, to evaluate higher
interaction vertices, we will look for an alternative calculation
method that avoids the computation of finite conformal
transformations. These methods known as conservation laws
\cite{Arroyo:2011zt,DeSmet:2001af,Rastelli:2000iu} already exist,
and we will need to implement them for our purposes.

\section*{Acknowledgements}
I would like to thank Ted Erler, Isao Kishimoto, and Michael
Kroyter for useful discussions. This work is supported by CNPq
grant 303073/2012-8.

\appendix
\setcounter{equation}{0}
\def\thesection{\Alph{section}}
\renewcommand{\theequation}{\Alph{section}.\arabic{equation}}

\section{Details on the level truncation evaluation of the vacuum energy}
We first consider some of the ingredients for the calculation of
the vacuum energy, then show a particular example in detail. First
of all, recall that for the cubic superstring field theory, the
inner product appearing in the evaluation of the vaccum energy is
the standard BPZ inner product with the difference that we must
insert the operator $Y_{-2}$ at the open string midpoint. The
operator $Y_{-2}$ can be written as the product of two inverse
picture changing operators $Y_{-2}=Y(i)Y(-i)$, where
$Y(z)=-\partial \xi e^{-2 \phi} c(z)$.

A general inner product of two vertex operators $\langle A , B
\rangle$ is given by the evaluation of the following correlator
\begin{eqnarray}
\label{inerx1} \langle A , B \rangle = \langle Y_{-2}
\big(\mathcal{I} \circ A(0)\big) B(0)\rangle,
\end{eqnarray}
where $\mathcal{I}(z)=-1/z$, is the usual conformal transformation
that defines the BPZ inner product. This definition, together with
the identity
\begin{eqnarray}
\label{corresuper002} &&\langle
\prod_{i=1}^{n}\xi(x_i)\prod_{j=1}^{n}\eta(y_j)\prod_{k=1}^{m}b(u_k)\prod_{l=1}^{m+3}c(v_l)\prod_{s=1}^{p}e^{q_s
\phi(z_s)}\rangle \nonumber \\
&=&
-\prod_{i<i'}(x_i-x_{i'})\prod_{j<j'}(y_j-y_{j'})\prod_{i,j}(x_i-y_j)^{-1}\prod_{k<k'}(u_k-u_{k'})\prod_{l<l'}(v_l-v_{l'})\prod_{k,l}(u_k-v_l)^{-1}
\nonumber \\
&&\times\prod_{s<s'}(z_s-z_{s'})^{-q_s q_{s'}},
\end{eqnarray}
and the expression for the bosonized representation of the
superconformal ghost $\gamma = \eta e^{\phi}$, allows us to
compute the relevant terms which appear in the level truncation
analysis of the vacuum energy.

For instance, employing the above identity (\ref{corresuper002}),
we have obtained some correlation functions that are very useful
in the evaluation of the vacuum energy presented in subsection
3.2, let us list these correlators
\begin{eqnarray}
\label{formu1} && \;\; \langle Y_{-2} c(x) \gamma(y) \gamma(z)
\rangle =
\frac{1}{2}(1+x^2)(1+y z),  \\
&& \;\; \langle Y_{-2} T(w)c(x) \gamma(y) \gamma(z) \rangle
\nonumber
\\&=& \frac{x^2-2 w x-1}{(1+w^2)(w-x)^2} \langle Y_{-2} c(x) \gamma(y)
\gamma(z) \rangle + \frac{1+x^2}{(1+w^2)(w-x)} \langle Y_{-2}
\partial_x c(x) \gamma(y) \gamma(z) \rangle \nonumber \\
&+& \frac{y^2-2 w y-1}{2(1+w^2)(w-y)^2} \langle Y_{-2} c(x)
\gamma(y) \gamma(z) \rangle + \frac{1+y^2}{(1+w^2)(w-y)} \langle
Y_{-2}
c(x) \partial_y \gamma(y) \gamma(z) \rangle \nonumber \\
\label{formu2} &+& \frac{z^2-2 w z-1}{2(1+w^2)(w-z)^2} \langle
Y_{-2} c(x) \gamma(y) \gamma(z) \rangle +
\frac{1+z^2}{(1+w^2)(w-z)} \langle Y_{-2} c(x) \gamma(y)
\partial_z \gamma(z) \rangle. \;\;\;\;\;\;\;
\end{eqnarray}

In general a correlation function of the form $\langle T(w)
\Phi_1(x) \Phi_2(y) \cdots \rangle$, where $\Phi_i$ is a primary
field with conformal weight $h_i$, can be computed using the
following expression
\begin{eqnarray}
&& \langle Y_{-2} T(w) \Phi_1(x) \Phi_2(y) \cdots \rangle
\nonumber
\\ &=& \frac{h_1(1+2 w x-x^2)}{(1+w^2)(w-x)^2} \langle  Y_{-2} \Phi_1(x)
\Phi_2(y) \cdots \rangle + \frac{1+x^2}{(1+w^2)(w-x)} \langle
Y_{-2}
\partial_x \Phi_1(x) \Phi_2(y) \cdots \rangle \nonumber \\
&+& \frac{h_2(1+2 w y-y^2)}{(1+w^2)(w-y)^2} \langle  Y_{-2}
 \Phi_1(x) \Phi_2(y) \cdots \rangle + \frac{1+y^2}{(1+w^2)(w-y)}
\langle  Y_{-2} \Phi_1(x) \partial_y \Phi_2(y) \cdots \rangle \nonumber \\
\label{formu3} &+& \cdots .
\end{eqnarray}

As a pedagogical illustration, we are going to show some steps in
the computation of the normalized value of the vacuum energy for a
string field expanded up to level two states. The expression for
this string field is given in equation (\ref{PsiY1}), and it can
be written using the corresponding vertex operators as follows
\begin{eqnarray}
\label{formu4} \Psi^{(1)}_\lambda= t(\lambda) c +
\frac{v(\lambda)}{2} \partial^2 c + w(\lambda) T c.
\end{eqnarray}
Plugging the string field (\ref{formu4}) into the definition of
the normalized value of the vacuum energy (\ref{energydependz}),
we obtain
\begin{align}
E_\lambda(z) = \frac{\pi^2}{3} \Big[ &
\frac{t^2(\lambda)}{z^2}\langle c,Qc\rangle + \frac{v^2(\lambda)
z^2}{4}\langle
\partial^2 c,Q\partial^2 c\rangle + w^2(\lambda) z^2 \langle
 T c,Q T c\rangle \nonumber \\ \label{formu5} & +  t(\lambda) v(\lambda) \langle
\partial^2 c,Q c\rangle + 2t(\lambda)w(\lambda) \langle
 T c,Q c\rangle + v(\lambda) w(\lambda) z^2 \langle
Tc,Q\partial^2 c\rangle \Big].
\end{align}

All correlation functions appearing in equation (\ref{formu5}) can
be evaluated using the correlators (\ref{formu1})-(\ref{formu3}).
For instance, let us explicitly compute the correlator $\langle
 T c,Q c\rangle$, since $Qc = c \partial c - \gamma^2 $, the
 non-vanishing contribution in the correlator $\langle
 T c,Q c\rangle$ is given by
\begin{eqnarray}
\label{formu6} \langle
 T c,Q c\rangle = - \langle
 T c,\gamma^2 \rangle.
\end{eqnarray}
Employing the definition of the BPZ inner product (\ref{inerx1}),
we can evaluate the above correlator by using a parameter
$\epsilon$ eventually taken to zero
\begin{eqnarray}
\label{formu7} \langle
 T c,Q c\rangle = - \lim_{\epsilon \rightarrow 0}\langle Y_{-2}
 \mathcal{I} \circ \big(T(\epsilon) c(\epsilon)\big)\,\gamma^2 (\epsilon)
 \rangle,
\end{eqnarray}
performing the conformal transformation of the vertex operator
$Tc$ and using equation (\ref{formu2}), we obtain
\begin{eqnarray}
\label{formu8} \lim_{\epsilon \rightarrow 0}\langle Y_{-2}
 \mathcal{I} \circ \big(T(\epsilon) c(\epsilon)\big) \, \gamma^2 (\epsilon)
 \rangle = 0 .
\end{eqnarray}
By the same procedure, we can compute all the remaining
correlators appearing in equation (\ref{formu5}). Therefore at the
end we arrive to an expression for the vacuum energy corresponding
to a string field truncated up to level two states
\begin{align}
\label{formu9} E_\lambda(z) = \frac{\pi^2}{3} \Big[ & -
\frac{t^2(\lambda)}{2 z^2} - \frac{v^2(\lambda) z^2}{2} + 2
w^2(\lambda) z^2  - t(\lambda) v(\lambda) + v(\lambda) w(\lambda)
z^2 \Big].
\end{align}


\end{document}